\documentclass[12pt, referee]{aastex}
\usepackage[dvips]{color}
\usepackage[11pt]{moresize}
\usepackage{lineno}

\newcommand{\bl}{\textcolor[rgb]{0,0,0.0}}

\def\gtsima{$\;\buildrel > \over \sim \;$}
\def\simgt{\lower.5ex \hbox{\gtsima}}
\def\ltsima{$\;\buildrel < \over \sim \;$}
\def\simlt{\lower.5ex \hbox{\ltsima}}

\begin{document}

\title{The chemistry of chlorine-bearing species in the 
diffuse interstellar medium, and new SOFIA/GREAT$^*$ observations of HCl$^+$}

\author{David A. Neufeld\altaffilmark{1}, Helmut Wiesemeyer\altaffilmark{2},
Mark J.\ Wolfire\altaffilmark{3}, Arshia Jacob\altaffilmark{2}, Christof Buchbender\altaffilmark{4},
Maryvonne Gerin\altaffilmark{5}, Harshal Gupta\altaffilmark{6}, Rolf G\"usten\altaffilmark{2}, and 
Peter Schilke\altaffilmark{4}}

\altaffiltext{1}{Department of Physics \& Astronomy, Johns Hopkins University, Baltimore, MD 21218, USA}
\altaffiltext{2}{Max-Planck-Institut f\"ur Radioastronomie, Auf dem H\"ugel 69, 53121 Bonn, Germany}
\altaffiltext{3}{Department of Astronomy, University of Maryland, College Park, MD 20742, USA}
\altaffiltext{4}{I. Physikalisches Institut, Universit\"at zu K\"oln, 50937 K\"oln, Germany}
\altaffiltext{5}{LERMA, Observatoire de Paris, PSL Research University, CNRS, Sorbonne Universit\'es, 75014 Paris, France}
\altaffiltext{6}{Division of Astronomical Sciences, National Science Foundation, Alexandria, VA 22314, USA}

\altaffiltext{*}{GREAT is a development by the MPI f\"ur Radioastronomie and the KOSMA/Universit\"at zu K\"oln, 
in cooperation with the MPI f\"ur Sonnensystemforschung and the DLR Institut f\"ur Planetenforschung.}
\begin{abstract}

We have revisited the chemistry of chlorine-bearing species in the 
diffuse interstellar medium with new observations
of the HCl$^+$ molecular ion and new astrochemical models.
Using the GREAT instrument on board SOFIA, we observed
the $^2\Pi_{3/2}\, J = 5/2 – 3/2$ transition of HCl$^+$ near 1444~GHz 
toward the bright THz continuum source W49N.  We detected absorption by diffuse
foreground gas unassociated with the background source, and were able to
thereby measure the distribution of HCl$^+$ along the sight-line.  We interpreted
the observational data using an updated version of an astrochemical
model used previously in a theoretical study of Cl-bearing interstellar molecules.  The 
abundance of HCl$^+$ was found to be almost constant relative to the related H$_2$Cl$^+$
ion, but the observed $n({\rm H_2Cl^+})/n({\rm HCl^+})$ abundance ratio exceeds the predictions
of our astrochemical model by an order-of-magnitude.  This discrepancy 
suggests that the rate of the primary destruction process for ${\rm H_2Cl^+}$, dissociative
recombination, has been significantly overestimated. 
For ${\rm HCl^+}$, the model predictions can provide a satisfactory fit to the observed 
column densities along the W49N sight-line while simultaneously accounting for
the ${\rm OH^+}$ and ${\rm H_2O^+}$ column densities.

\end{abstract}


\section{Introduction}

Hydride molecules, consisting of a single heavy element atom with one or
more hydrogen atoms, have been studied extensively in the interstellar
medium (ISM).  Such molecules are valuable probes of the environment
in which they are found; collectively, they provide unique information about 
the density of cosmic rays, the prevalence of shocks, and the dissipation of interstellar 
turbulence (Gerin et al.\ 2016 and references therein).
To date, 19 interstellar hydrides have been detected in diffuse interstellar 
gas clouds containing the elements C, N, O, F, S, Cl, or Ar.  For most of these 
elements, their interstellar hydrides are only a trace constituent in diffuse atomic 
and molecular clouds; 
here, the dominant reservoirs are atoms or atomic ions,
and hydrides typically account for only $\sim 0.01\%$ the overall gas-phase inventory 
of a given element (Gerin et al.\ 2016; their Table 2).  
The two exceptions are the halogen elements, F and Cl.
The very stable hydrogen fluoride molecule typically accounts for $\sim 40\%$ of the gas-phase
fluorine in diffuse molecular clouds, and the molecular ions HCl$^+$ and 
H$_2$Cl$^+$ may account for several percent of the gas-phase chlorine.

The great tendency of fluorine and chlorine to form hydride molecules in
the diffuse ISM is a consequence of thermochemistry (e.g. Neufeld \& Wolfire 2009;
hereafter NW09): they are the only
two elements that can react exothermically with molecular hydrogen
when in their primary ionization state in diffuse atomic clouds (i.e.\ as
F and Cl$^+$, respectively).  While the observed abundance of HF (e.g. 
Sonnentrucker et al.\ 2015) is
well accounted for by astrochemical models (NW09),
such models have tended to underpredict the observed abundances of 
H$_2$Cl$^+$ (Lis et al.\ 2010; Neufeld et al.\ 2012; 2015) and
HCl$^+$ (De Luca et al.\ 2012; hereafter DL12).

In the present paper, we address the puzzle of the interstellar
HCl$^+$ and H$_2$Cl$^+$ abundances with additional observational
and theoretical studies.  The only previous detections of interstellar
HCl$^+$ were obtained by DL12
using the HIFI instrument on the {\it Herschel} Space Telescope. 
DL12 detected the 1444~GHz $^2\Pi_{3/2}\,  J = 5/2 – 3/2$ transition 
of HCl$^+$ in absorption toward the bright THz continuum sources
W49N and W31C.  The Doppler velocities of the detected absorption
indicated that it arose in diffuse foreground gas clouds along the 
sight-lines to -- but physically unassociated with -- these background 
continuum sources.  Although these HCl$^+$ detections were unequivocal, the extraction of
reliable column densities was made more difficult by the complex
hyperfine structure of HCl$^+$ and by the technical challenges
entailed by operating at the required frequency near the very bottom of 
HIFI Band 6.  Several spectra had to be discarded due to insufficient
mixer pumping, and instrumental drifts -- to which the hot electron bolometer 
mixers were prone -- led to strong standing waves that had to be removed.  
As a result of these difficulties, the observed HCl$^+$ spectra had to be fit by
using H$_2$Cl$^+$ as a template for the distribution of HCl$^+$
in velocity space along the line-of-sight.  
This procedure yielded the total HCl$^+$ column density along the sight-lines 
to W49N and W31C but did not permit a determination of the column densities
over specific velocity intervals within the absorbing gas.  The two sources showed
almost identical absorbing column densities of HCl$^+$, and very similar 
HCl$^+$ to H$_2$Cl$^+$ abundance ratios.  

With the goal of improving the observational
data available for the study of interstellar HCl$^+$, we have reobserved its
$^2\Pi_{3/2}\, J = 5/2 – 3/2$ ground-state transition toward W49N using the 
GREAT instrument (Heyminck et al.\ 2012) on SOFIA .  The new observations and data reduction
are described in Section 2 below, and the results are discussed in Section 3.
We have compared the measured column densities of HCl$^+$ and 
H$_2$Cl$^+$ (the latter observed previously by Neufeld et al.\ 2015; hereafter N15) with
the predictions of a grid of astrochemical models.  These predictions are based on 
the NW09 model, which we have updated to account for several laboratory
experiments or theoretical calculations performed in the last
decade. These developments required modifications to the adopted rates of several significant
processes that form or destroy HCl$^+$ and H$_2$Cl$^+$.  The astrochemical model
is described in Section 4.  In Section 5, we discuss the comparison between the
astronomical observations and the predictions of the astrochemical model.  A brief
summary follows in Section 6.

\section{New observations and data reduction}

The observations of W49N were performed on 2016 May 18 UT with the GREAT receiver
on SOFIA.  The $^2\Pi_{3/2}\, J = 5/2 – 3/2$ transition of HCl$^+$ was observed in the
lower sideband of the L1 receiver.  At the observing frequency of 1444~GHz, the 
half power beam width of the telescope was $\sim 20^{\prime\prime}$.
The observations were performed with the AFFTS backend, which provides 16384 spectral 
channels with a spacing of 244.1 kHz.  
The dual beam switch mode was used to carry out the observation, with a chopper frequency of
2.5~Hz and the reference positions located 160$^{\prime\prime}$ on 
either side of the source along an east-west axis.

Using an independent fit to the
dry and the wet content of the atmospheric emission, 
the raw data were calibrated to the $T_A^*$ (“forward beam
brightness temperature”) scale.  The assumed forward efficiency was 0.97 
and the uncertainty in the flux calibration is estimated to be
$\sim 20\%$ (Heyminck et al. 2012).
Additional data reduction was
performed using CLASS \footnote{Continuum~and~Line~Analysis~Single-dish~Software 
\phantom{0000000000000000000000000000}
(http://www.iram.fr/IRAMFR/GILDAS)}
to rebin the data into 20-channel bins (each corresponding to a velocity width of
1.0$\rm  \, km\,s^{-1}$), remove a linear baseline, perform the  
conversion to a single sideband spectrum, and finally obtain a transmission spectrum
by dividing the single sideband antenna temperature, $T_A^*({\rm SSB})$, by the continuum 
temperature, \bl{$T_{\rm cont} = 6.2 \, \rm K$}.  The latter corresponds to a 
main-beam brightness temperature of 9.1 K, given
a main beam efficiency of 0.66 as measured on Jupiter.

\section{Results}

\begin{figure}
\includegraphics[scale=0.7]{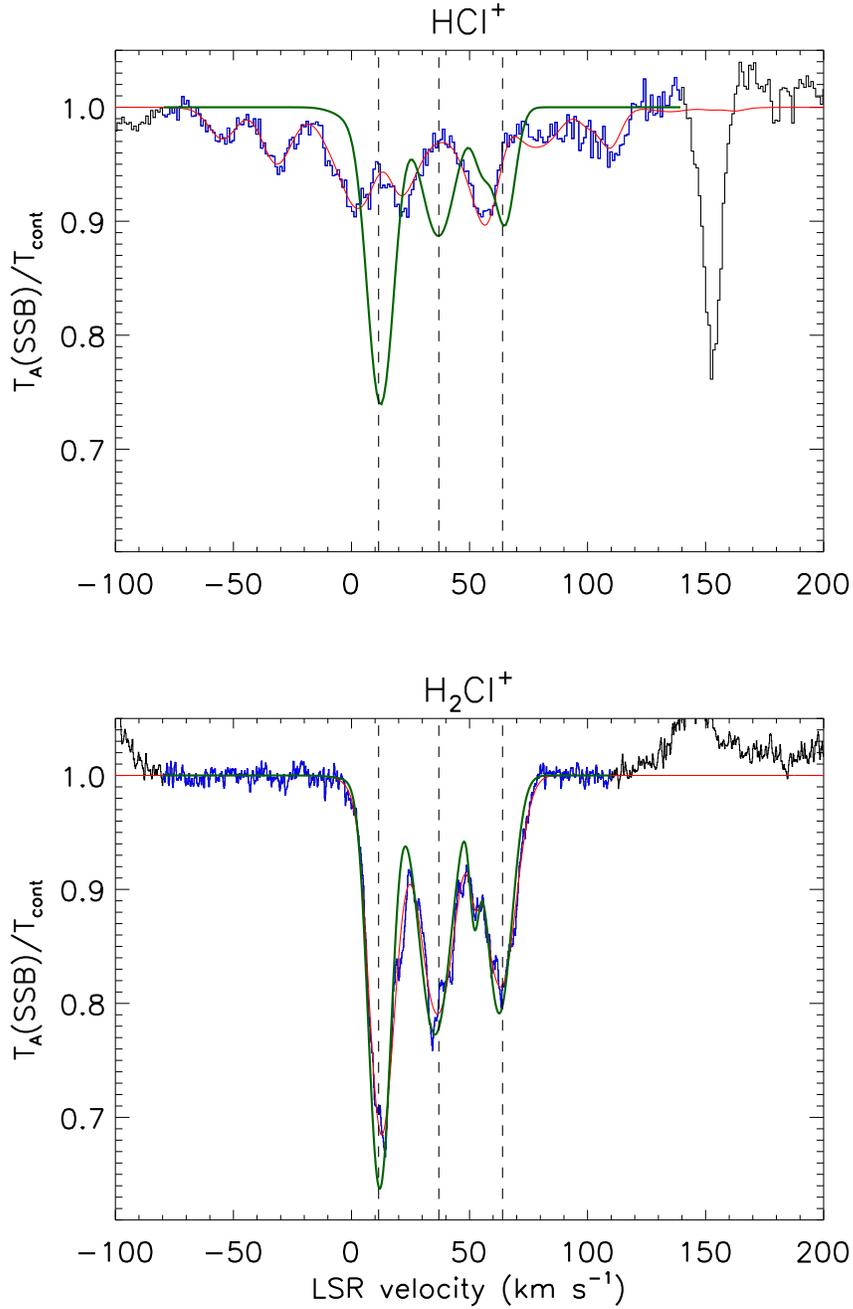}
\caption{Single-sideband W49N absorption spectra for HCl$^+$ (SOFIA/GREAT; top panel; present work)
and H$_2$Cl$^+$ ({\it Herschel}/HIFI; bottom panel; N15).  Histograms: observed spectra 
normalized relative to the continuum flux.  Green curves: hyperfine-deconvolved spectra
obtained for the blue portions of the histograms.  
Red curves: convolution of the green curves with the hyperfine structure.  The fitting procedure
(see the text for details) optimizes the agreement between the blue portion of the 
histogram and the red curve.} 
\end{figure}

The HCl$^+$ spectrum observed toward W49N is presented
in the upper panel of Figure 1 (histogram),
where the transmission, $T_A^*({\rm SSB})/T_{\rm cont}^*({\rm SSB})$ is shown as a function 
of LSR velocity, $v_{\rm LSR}$; the latter is shown for an assumed line rest frequency of
1444.3~GHz.  In reality, the effects of lambda-doubling and the nuclear
hyperfine interaction split the  $^2\Pi_{3/2}\,  J = 5/2 – 3/2$ transition into 
18 transitions spanning the 1443.643 -- 1444.635~GHz frequency range 
(DL12, their Table 1).  Thus, the observed spectrum (covering only part of
this frequency range) represents a convolution between the velocity structure
of the absorbing material and the hyperfine structure of the absorption.
To deconvolve the observed spectrum and recover the intrinsic velocity 
structure of the absorbing material, we represented the column density
per unit velocity interval as the sum
of $N$ Gaussians, each with an adjustable height, width and centroid position.
The $3N$ resulting parameters were then varied to optimize the agreement
between the observed spectrum and the convolution of the velocity structure
with the hyperfine structure.  The optimization was
performed using the Levenberg-Marquardt algorithm as implemented in the
IDL routine {\tt mpfit.pro} (Markwardt 2009).  Five Gaussian components ($N=5$) 
were found sufficient
to provide an excellent fit to the data (red curve in Figure 1).  The 
``hyperfine-convolved" spectrum, shown in green, represents the absorption 
spectrum that would have resulted in the absence of hyperfine structure. 

In the deconvolution procedure described above, 
only the $[-80,+140]\,{\rm km\,s^{-1}}$
velocity interval was fit (blue portion of the histogram shown
in Figure 1).  A strong para-NH$_2$ absorption feature appearing at 
$+ 155 \, \rm km\,s^{-1}$ was thereby excluded.
Gaussian centroid velocities were permitted to range over the 
$[0,+65]\,{\rm km\,s^{-1}}$ velocity interval for which other
molecules are known to absorb along the sight-line to W49N.
While the deconvolution procedure should preserve the
equivalent width of optically-thin lines, the 
equivalent width of the deconvolved spectrum
significantly exceeds that of the observed spectrum shown in Figure 1.  
The explanation for this discrepancy is that much of the 
expected absorption in the convolved spectrum lies outside
the frequency range covered by the observations.

For comparison, the histogram
in the lower panel of Figure 1 shows H$_2$Cl$^+$ spectrum obtained
by Neufeld et al.\ (2012), together with the deconvolved spectrum (green)
and the fit to the observed spectrum (red).  The two molecular
ions exhibit remarkably similar distributions in $v_{\rm LSR}$,
as is demonstrated in Figure 2 (top panel) where the column density
per unit velocity interval, $dN/dv$, is shown for each species.
\bl{Here, we adopted the line frequencies and 
strengths given by DL12 for HCl$^+$ and 
the Cologne Database for Molecular Spectroscopy 
(CDMS; M\"uller et al.\ 2001, 2005; Endres et al.\ 2016)
for H$_2$Cl$^+$, line frequencies for the latter being based
on laboratory spectroscopy reported by Araki et al.\ (2001).
The line strengths are for assumed dipole moments of 1.75~Debye for HCl$^+$ 
(calculation of Cheng et al.\ 2007) and 1.89~Debye for H$_2$Cl$^+$ 
(unpublished calculation by H. S. P. M\"uller 2008)}.

We have computed the HCl$^+$ column density for seven
velocity intervals that have been defined in previous 
studies of W49N (e.g.\ Indriolo et al.\ 2015 and references therein).
The results are listed in Table 1, along with the column densities
of atomic hydrogen (Winkel et al.\ 2017).   We also list 
the average abundances of HCl$^+$ and H$_2$Cl$^+$ (N15)
in each velocity interval, relative to atomic hydrogen, along with those of
two other molecular ions observed with {\it Herschel}
(Indriolo et al.\ 2015): $\rm OH^+$ and $\rm H_2O^+$.
Finally, we compute the values of two column density ratios
that are expected to probe the molecular hydrogen fraction
$N({\rm H_2Cl^+})/ N({\rm HCl^+})$ and $N({\rm H_2O^+})/ N({\rm OH^+})$.
The results for $N({\rm HCl^+})/N({\rm H})$ and $N({\rm H_2Cl^+})/N({\rm H})$ 
are also represented graphically in Figure 2 (middle panel), as
are those for $N({\rm H_2Cl^+})/ N({\rm HCl^+})$ (bottom panel).

\begin{deluxetable}{lcccccc}
\tabletypesize \scriptsize
\tablewidth{0pt}
\tablecaption{Column densities and column density ratios observed toward W49N }

\tablehead{
Velocity interval
&$N({\rm HCl^+})$ 			
&$N({\rm H})$				
&$N({\rm HCl^+})/N({\rm H})$	
&$N( {\rm H_2Cl^+})/N({\rm H})$ 
&$N({\rm H_2Cl^+})/N({\rm HCl^+})$ 
&$N({\rm H_2O^+})/N({\rm OH^+})$ \\
$(\rm km\,s^{-1})$
&$(10^{13}\rm \,cm^{-2})$	
&$(10^{21}\rm \,cm^{-2})$  
&$\times 10^9$
&$\times 10^9$
}
\startdata
 $[ 1, 10]$ & $  1.15  \pm   0.34 $ & $  4.44  \pm   0.18 $ & $  2.59  \pm   0.78 $ & $  2.84  \pm   0.85 $ & $  1.10  \pm   0.33 $ & $ 0.191  \pm  0.025$\\
 $[10, 17]$ & $  1.65  \pm   0.49 $ & $  3.58  \pm   0.13 $ & $  4.61  \pm   1.38 $ & $  6.30  \pm   1.89 $ & $  1.37  \pm   0.41 $ & $ 0.347  \pm  0.029$\\
 $[17, 25]$ & $  0.68  \pm   0.20 $ & $  2.34  \pm   0.08 $ & $  2.90  \pm   0.87 $ & $  3.43  \pm   1.03 $ & $  1.18  \pm   0.36 $ & $ 0.104  \pm  0.011$\\
 $[25, 43]$ & $  1.42  \pm   0.43 $ & $  5.58  \pm   0.17 $ & $  2.55  \pm   0.77 $ & $  5.02  \pm   1.51 $ & $  1.97  \pm   0.59 $ & $ 0.190  \pm  0.015$\\
 $[43, 51]$ & $  0.36  \pm   0.11 $ & $  1.67  \pm   0.05 $ & $  2.13  \pm   0.64 $ & $  3.54  \pm   1.06 $ & $  1.66  \pm   0.50 $ & $ 0.134  \pm  0.015$\\
 $[51, 66]$ & $  0.97  \pm   0.29 $ & $  6.15  \pm   0.23 $ & $  1.58  \pm   0.47 $ & $  3.46  \pm   1.04 $ & $  2.20  \pm   0.66 $ & $ 0.143  \pm  0.014$\\
 $[66, 80]$ & $  0.40  \pm   0.12 $ & $  2.89  \pm   0.11 $ & $  1.38  \pm   0.42 $ & $  2.24  \pm   0.67 $ & $  1.62  \pm   0.49 $ & $ 0.127  \pm  0.021$\\

\\
\multicolumn{7}{l}{$^a$ Column densities considered unreliable for this velocity interval,
which lies close to the systemic velocity of the background source}\\
\\
\enddata
\end{deluxetable}

The results shown for $N({\rm H_2Cl^+})$ and $N({\rm HCl^+})$, both  
in Figure 2 and Table 1, are estimates for the total column densities
including both stable isotopologs and both spin-isomers for ${\rm H_2Cl^+}$.
Here, we inferred those column densities from the observations 
of $\rm H^{35}Cl^+$ and ortho-${\rm H_2^{35}Cl^+}$ by assuming
and isotopic ratio $\rm ^{35}Cl/^{37}Cl = 3.1$ with no fractionation
and an ortho-to-para ratio of 3 for ${\rm H_2Cl^+}$ (N15).

The overall shape of the HCl$^+$ spectrum shown in Figure 1 agrees well with
that presented by DL12 (their Figure 1), except in regions where the DL12 
results were considered unreliable (shaded regions in their Figure 3) and were 
excluded from their fit.
One notable difference, however, is that all the absorption lines measured in the present study
are $\sim 30 - 40\%$ less deep than those in the DL21 spectrum.
This is seen most clearly in the narrow para-$\rm NH_2$ feature: in Figure 1 of the present study,
the interstellar transmission at line center is 78$\%$, whereas in the DR12 spectrum it
is only $\sim 66\%$.  Both the DL12 {\it Herschel}/HIFI spectrum and the GREAT spectrum
obtained with SOFIA were acquired with dual sideband receivers, and we speculate that 
uncertainties in the sideband gain ratios (SBR) might be cause of these discrepancies.
\footnote{\bl{In the case of {\it Herschel}/HIFI, the SBR has been examined carefully by 
Kester et al.\ (2017), but the result may be affected by the instrumental challenges 
in working at the HCl$^+$ frequency (described above in Section 1).  In the case of
SOFIA/GREAT, uncertainties are introduced by the presence of an atmospheric
ozone line with wings that may not be described entirely accurately by atmospheric models.}}
Accordingly, in Table 1, we estimate the uncertainties in the ${\rm HCl^+}$ column densities
as $\pm 30\%$.

\begin{figure}
\includegraphics[scale=0.7]{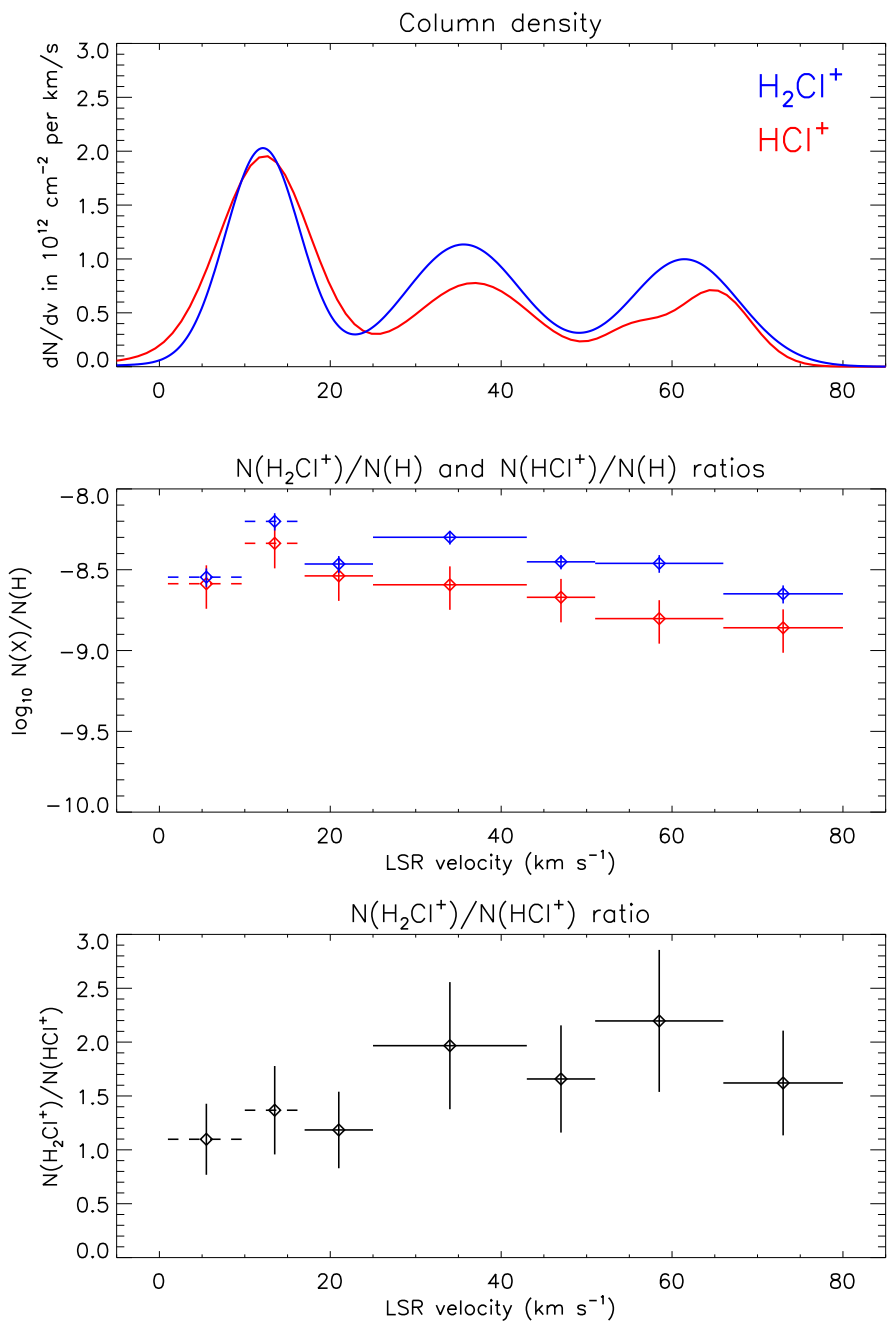}
\caption{Top panel: column density per unit velocity interval, $dN/dv$, for HCl$^+$ (red)
and H$_2$Cl$^+$ (blue).  Middle panel: average abundances of HCl$^+$ (red)
and H$_2$Cl$^+$ (blue) relative to atomic hydrogen for 7 velocity intervals defined by 
Indriolo et al.\ (2015).  
Bottom panel: $N({\rm H_2Cl^+})/N({\rm HCl^+})$ for each velocity interval.
The data represented by dotted lines in the middle and bottom panels indicate values
obtained for material close to the systemic velocity of the background source where
the results may be unreliable.} 
\end{figure}

\vfill \eject
\section{Model}

To interpret our observations of ${\rm HCl^+}$ and ${\rm H_2Cl^+}$, we have constructed 
a grid of interstellar cloud models similar to that described by Neufeld 
\& Wolfire (2017; hereafter NW17).  In NW17, we obtained predictions for the equilibrium 
column  densities of ${\rm OH^+}$ and ${\rm H_2O^+}$ and other species 
for a grid of 1440 isochoric cloud 
models.   Our grid covers 9 values of the primary cosmic ray ionization rate per H 
nucleus, $\zeta_p({\rm H}) =$ 0.006, 0.02, 0.06, 0.2, 0.6, 2.0, 6.0, 20, and 
$60 \times 10^{-16}\, \rm s^{-1}$; 10 values of the interstellar radiation field, $\chi_{UV}$, 
defined here as the ratio of the UV energy density to the average interstellar
value estimated by Draine (1978): 0.05, 0.1, 0.2, 0.3, 0.5, 1, 2, 3, 5, and
10; and 16 values of the total visual extinction through the
cloud, $A_V({\rm tot})$: 0.0003, 0.001, 0.003, 0.01, 0.03, 0.1, 0.2,
0.3, 0.5, 0.8, 1.0, 1.5, 2.0, 3.0, 5.0, and 8.0 mag.  
The grid of models was computed for
a single density of H nuclei, $n_{\rm H} 
= n({\rm H})+2n({\rm H}_2)= 50\, \rm cm^{-3}$.  
However, the
cloud properties can be predicted for other values of $n_{\rm H}$ by
means of a simple scaling because the cloud properties are
completely determined by $\zeta_p({\rm H})/n_{50}$, $\chi_{UV}/n_{50}$,
and $A_V({\rm tot})$, where $n_{50} \equiv n_{\rm H}/(50\, \rm cm^{-3})$.
We will refer to $\zeta_p({\rm H})/n_{50}$ and $\chi_{UV}/n_{50}$ as
the ``environmental parameters," because they reflect the environment
(pressure, cosmic ray density, UV energy density) in which a cloud is located.
The third parameter, $A_V({\rm tot})$, is related to the cloud column density.

Details of the diffuse cloud model, which solves for the equilibrium
physical and chemical
structure of a cloud with a slab geometry, have been given in N17, Hollenbach et al.\ 
(2012), and references therein: they will not be repeated here except 
in section 4.2 where minor updates to the model are enumerated. 
In the present study, our diffuse cloud model was supplemented by the inclusion
of additional processes governing the chemistry of chlorine-bearing species, leading to 
predictions for the column densities of ${\rm HCl^+}$ and ${\rm H_2Cl^+}$. 

\subsection{Chemical network for chlorine-bearing species}

Figure 3 represents the chemical network we adopted, with key reactions shown in red. 
The basic structure of this network is similar to that discussed by NW09, 
but several reaction rates have been updated to reflect recent laboratory
measurements or new theoretical calculations.   The rate coefficients or photorates 
that we adopted are listed in Table 2.  

\begin{figure}
\includegraphics[scale=0.8]{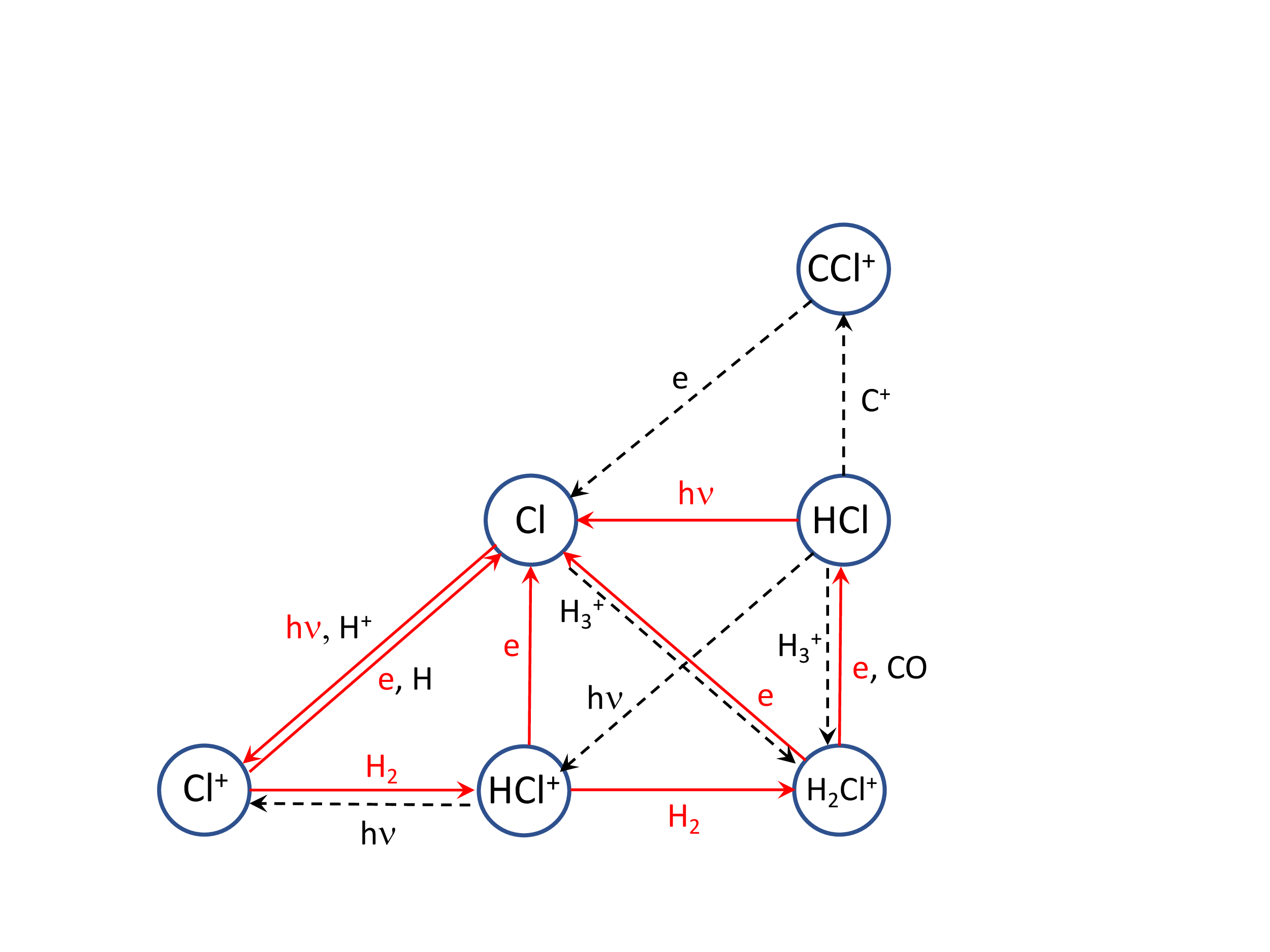}
\caption{Diagram representing the chemical network for chlorine-bearing species
in the ISM.  The most important processes in diffuse atomic and
molecular clouds are shown with
solid red arrows.} 
\end{figure}

\begin{deluxetable}{llll}
\tabletypesize \scriptsize
\tablewidth{0pt}
\tablecaption{Reaction list for Cl-bearing species}
\tablehead{&Reaction & Rate ($\rm s^{-1})$ or rate coefficient ($\rm cm^3\,s^{-1}$) & Reference$^a$ \\}
\startdata
(1) & $\rm Cl + H_2 \rightarrow HCl + H  \phantom{00}$ & $2.52 \times \rm 10^{-11}\,\exp\,(-2214\,K/{\it T})$ 
$\phantom{0000000000000} \rm ({\it T} \le 354\, K)$\\
&& $\rm 2.86 \times \rm 10^{-12}\,({\it T}/300\,K)^{1.72}\,\exp\,(-1544\,K/{\it T})$ 
$\,\, \rm ({\it T} \ge 354 \,K)$\\
(2) & $\rm HCl + H \rightarrow Cl + H_2  \phantom{00}$ & $1.49 \times \rm 10^{-11}\,\exp\,(-1763\,K/{\it T})$ 
$\phantom{0000000000000} \rm ({\it T} \le 354\, K)$\\
&& $\rm 1.69 \times \rm 10^{-12}\,({\it T}/300\,K)^{1.72}\,\exp\,(-1093\,K/{\it T})$ 
$\,\, \rm ({\it T} \ge 354 \,K)$\\
(3) & $\rm  C^+ + HCl \rightarrow CCl^+ + H  $ & $1.84 \times 10^{-9} \, (T/300 \rm \, K)^{-0.202}$\\
(4) & $\rm  H_3^+ + HCl \rightarrow H_2 + H_2Cl^+ $ & $6.35 \times 10^{-9} \, (T/300 \rm \, K)^{-0.202}$\\
(5$^*$) & $\rm  Cl^+ + H_2 \rightarrow HCl^+ + H $ & $1.0 \times \rm 10^{-9} $ \\
(6$^*$) & $\rm  HCl^+ + H_2 \rightarrow H_2Cl^+ + H $ & $1.3 \times \rm 10^{-9} $\\
(7) & $\rm CCl^+ + e \rightarrow C + Cl $ & $2.0 \times 10^{-7} \, (T/300 \rm \, K)^{-0.5}$\\
(8$^*$) & $\rm HCl^+ + e \rightarrow H + Cl $ & Fitting formula given by N13 &  N13 \\
(9$^*$) & $\rm H_2Cl^+ + e \rightarrow products $ & $4.3 \times 10^{-7}\, (T/300 \rm\, K)^{-0.5}$ & N18\\
(10) & $\rm H_2Cl^+ + CO \rightarrow HCl + HCO^+$ & $\rm 7.8 \times 10^{-10}\,$\\
(11) & $\rm H_2Cl^+ + H_2O \rightarrow HCl + H_3O^+$ & $3.7 \times 10^{-9}\, (T/300 \rm\, K)^{-0.194}$\\
(12) & $\rm Cl + H_3^+ \rightarrow H_2Cl^+ + H$ & $1.0 \times \rm 10^{-9} $\\
(13) & $\rm HCl + h\nu \rightarrow H + Cl $ & $\,1.7 \times 10^{-9}\,\chi_{UV} [\,{1 \over 2}E_2(2.02 A_V) + {1 \over 2}E_2(2.02 [A_{V,{\rm tot}}-A_V])] $ \\
(14) & $\rm HCl + h\nu \rightarrow HCl^+ + e$ & $\,4.5 \times 10^{-11}\,\chi_{UV} [\,{1 \over 2}E_2(3.18 A_V)f_1(N_a) 
+ {1 \over 2}E_2(3.18 [A_{V,{\rm tot}}-A_V])f_1(N_b)]$ & H17 $^b$\\
(15$^*$) & $\rm Cl + h\nu \rightarrow Cl^+ + e$ & $\,4.7 \times 10^{-11}\,\chi_{UV} [\,{1 \over 2}E_2(3.21 A_V)f_2(N_a) 
+ {1 \over 2}E_2(3.21 [A_{V,{\rm tot}}-A_V])f_2(N_b)]$ & H17 $^b$\\
(16) & $\rm HCl^+ + h\nu \rightarrow H + Cl^+ $ & $\,1.1 \times 10^{-10}\,\chi_{UV} [\,{1 \over 2}E_2(2.12 A_V) 
+ {1 \over 2}E_2(2.12 [A_{V,{\rm tot}}-A_V])] $ & H17\\
(17$^*$) & $\rm Cl^+ + e \rightarrow Cl + h\nu $ & $1.34 \times 10^{-11}\, (T/300 \rm \, K)^{-0.738}$\\ 
(18) & $\rm Cl^+ + H \rightarrow Cl + H^+$   & $6.2 \times \rm 10^{-11}\, (T/300 \rm \, K)^{0.79}\,\exp\,(-6920\,K/{\it T})$ \\
(19) & $\rm Cl + H^+ \rightarrow Cl^+ + H$   & $9.3 \times \rm 10^{-11}\, (T/300 \rm \, K)^{0.73}\,\exp\,(-232\,K/{\it T})$ \\
(20) & $\rm HCl + H^+ \rightarrow HCl^+ + H$ & $3.3 \times \rm 10^{-9}\, (T/300 \rm \, K)^{1.00}$ \\
\hline
\multicolumn{4}{l}{$^*$ Important reaction in diffuse interstellar clouds} \\
\multicolumn{4}{l}{$^a$ References are given where the rate adopted differs from that in NW09} \\
\multicolumn{4}{l}{$\,\,\,\,\,$N13 = Novotn{\'y} et al.\ (2013) and see the text; N18 = Novotn{\'y} et al.\ (2018), H17 = Heays et al.\ (2017)} \\
\multicolumn{4}{l}{$^b$ $N_a$ and $N_b=N({\rm H_2})-N_a$ are the H$_2$ column densities to the 
cloud surfaces.  Shielding by H$_2$ is accounted for (NW09) with} \\
\multicolumn{4}{l}{$\,\,\,\,\,$the functions $f_1(10^{21} N_{21}\,{\rm cm}^{-2})=\exp(-N_{21}/3.5) / (1 + N_{21}/0.32)^{1/2}$
or $f_2(10^{21} N_{21}\,{\rm cm}^{-2})=\exp(-N_{21}/4.28) / (1 + N_{21}/0.49)^{1/2}$ }
\enddata
\end{deluxetable}

Among the key reactions shown by red arrows in Figure 3, the dissociative recombination (DR) rates for HCl$^+$ and H$_2$Cl$^+$ are of
critical importance and have recently been the subject of extensive laboratory investigation.
For the case of DR of $\rm HCl^+$ (Table 2, Reaction 8), and in the absence of any 
laboratory data,  NW09 assumed the rate coefficient to be typical of those
for the DR of diatomic molecular ions and adopted a value
$k_{\rm DR}({\rm HCl^+}) = 2 \times 10^{-7} (T/{\rm 300\,K})^{-1/2} \rm cm^3 \,s^{-1}$. 
Over the past decade,
two laboratory determinations of this key reaction rate have been undertaken 
and are in good agreement with each other.
First, measurements performed by Novotn{\'y} et al.\ (2013; hereafter N13) at the TSR heavy-ion 
storage ring in Heidelberg used a merged beams configuration to derive DR rate coefficients
that were 1.5, 1.1, 0.64, 0.33, and 0.16 times as large as this generic
estimate at temperatures of 10, 30, 100, 300, and 1000 K, respectively.
Subsequently, Wiens et al.\ (2016; hereafter W16) measured the DR rate coefficient for several 
chlorine-bearing molecular ions, including $\rm HCl^+$,
under thermal conditions 
with the use of a flowing afterglow –- Langmuir probe apparatus; their
results for $\rm HCl^+$ DR were entirely consistent with those of N13 
to within the estimated uncertainties.  Accordingly, we adopted N13's
estimates of the rate coefficient of HCl$^+$ DR and used their fitting
formula to characterize its temperature dependence.

In the case of $\rm H_2Cl^+$ DR, the experimental picture is less clear.  For this
reaction, NW09 adopted a rate coefficient provided by an unpublished storage ring 
(CRYRING) experiment involving the $\rm D_2Cl^+$ isotopologue
(Geppert et al.\ 2009, private communication):
$k_{\rm DR}({\rm D_2Cl^+}) = 1.2 \times 10^{-7} (T/{\rm 300\,K})^{-0.85} \rm cm^3 \,s^{-1}$.
NW09 assumed the $\rm H_2Cl^+$ DR rate to be equal to that for $\rm D_2Cl^+$.
In the past decade, three additional experiments of relevance have been performed.
Kawaguchi et al.\ (2016) obtained an estimate of the total DR rate for $\rm H_2Cl^+$
using absorption spectroscopy in a pulsed discharge plasma.  Their value, obtained
at 209~K, was a factor 3 smaller than the CRYRING value for $\rm D_2Cl^+$.
This study was followed by the flowing afterglow investigation of W16, which
yielded a DR rate coefficient for $\rm D_2Cl^+$ at 300~K that was in excellent agreement
with the CRYRING value, and a DR rate coefficient for $\rm H_2Cl^+$ that was roughly
twice as large.  Interestingly, the temperature dependence inferred for
DR of both isotopologues over the 300 -- 500~K range 
was considerably stronger than in the CRYRING estimate with $k_{\rm DR} \propto T^{-1.4}$.
Most recently, another determination of the DR rate coefficient for $\rm D_2Cl^+$
has been obtained at the TSR heavy-ion storage ring (Novotn{\'y} et al.\ 2018; hereafter 
N18): $k_{\rm DR}({\rm D_2Cl^+}) = 4.3 \times 10^{-7} (T/{\rm 300\,K})^{-0.5} \rm cm^3 \,s^{-1}$.
\footnote{\bl{The expression for $k_{\rm DR}({\rm D_2Cl^+})$ given by N18 is most reliable 
at low temperatures $\simlt 20$~K, where the estimated uncertainty is $\sim 12\%$.  
At higher temperatures, the result becomes increasingly
dependent on an extrapolation of the DR cross-section to energies higher than that 
at which the measurements were made; the expression given assumes that the cross-section
is inversely proportional to the energy.}}  
At 300 K, this expression yields a value that is a factor $\sim 3$ larger than the results obtained for
$\rm D_2Cl^+$ by Geppert et al.\ (2009) and W16; and at 209 K, it is an
order-of-magnitude larger than the $\rm H_2Cl^+$ DR rate reported by Kawaguchi et al.\ (2016).
The ambiguous experimental picture discussed above is summarized nicely by Figure 3 in N18.

In our standard model, we have adopted the $\rm D_2Cl^+$ DR rate given by N18
(Table 2, Reaction 9) and assumed the $\rm H_2Cl^+$ DR rate to be identical.  
But we have also obtained column density predictions for HCl$^+$ and H$_2$Cl$^+$ with
an assumed $k_{\rm DR}({\rm H_2Cl^+})$ one-tenth this standard value.  As will be
discussed below, the smaller $\rm H_2Cl^+$ DR rate yields predicted $\rm H_2Cl^+$ column densities
that are an order-of-magnitude larger without significantly affecting
the predictions for $N({\rm HCl^+})$; it also provides a much better fit to the observed 
$\rm H_2Cl^+$ abundance, which is significantly underpredicted by the standard model.
As in previous astrophysical studies, we assume a branching 
fraction to HCl of 10$\%$ following DR of $\rm H_2Cl^+$.  We note, however, that
the predicted abundances of HCl$^+$ and H$_2$Cl$^+$ are essentially independent of 
the branching fraction assumed here (although the predicted HCl abundance scales 
linearly with that fraction).

One additional uncertainty applies to all the DR rates discussed above.  While all the 
experimental values apply to rotationally-warm molecular ions, HCl$^+$ and 
H$_2$Cl$^+$ are primarily present in their ground rotational states at the low 
densities of diffuse interstellar clouds.  Future experiments with the new cryogenic
storage ring (CSR) in Heidelberg will be needed to determine whether there is a 
significant dependence on the rotational state of the molecular ion.  This
has recently been found to be the case for DR of HeH$^+$ (Novotn{\'y} et al.\ 2019), raising
a significant caveat about the applicability of the laboratory DR measurements
for HCl$^+$ and H$_2$Cl$^+$ discussed above.

\subsection{Other minor updates to the diffuse cloud models}

Several additional changes to the NW17 diffuse cloud models have been implemented in
the present study, as described below. The combined effect of these changes on the model 
predictions is found to be minor.

\noindent(1) Wherever available, the photorates of Heays et al.\ (2017) were adopted,
with the $E_2(kA_{\rm V})$ attenuation factor appropriate to isotropic radiation and the
standard UV spectral shape given by Draine (1978).  

\noindent(2) The rate coefficient determined in the recent ion-trap experiment
of Kovalenko et al.\ (2018) was adopted for the key hydrogen atom abstraction
reaction\footnote{Here, we denote the reaction $\rm A + B \rightarrow C + D$ with the
notation A(B,C)D} $\rm O^+(H_2,H)OH^+$: $k=1.3 \times 10^{-9}\,\rm cm^3\,s^{-1}$.

\noindent(3) The rate coefficients determined in the recent ion-trap experiment
of Tran et al.\ (2018) were adopted for the key hydrogen atom abstraction
reactions $\rm OH^+(H_2,H)H_2O^+$ ($k=1.0 \times 10^{-9}\,\rm cm^3\,s^{-1}$)
and $\rm H_2O^+(H_2,H)H_3O^+$ ($k=0.97 \times 10^{-9}\,\rm cm^3\,s^{-1}$).

\noindent(4) We adopted the rate coefficients computed recently by Dagdigian et al.\
(2019) for the reactions $\rm C^+(OH,H)CO^+$ and $\rm C^+(OH,H^+)CO$.

\noindent(5) We adopted collisional rate coefficients computed recently by Lique et al.\ 
(2018; and private communication) 
for the excitation of fine structure states of atomic oxygen by H and H$_2$.


\noindent(6) We adopted the rate coefficient
measured by de Ruette et al.\ (2015) for the reaction $\rm O(H_3^+,H_2)OH^+$.

\noindent(7) We adopted the rate coefficients recommended by the KIDA
reaction list (Wakelam et al.\ 2015) for the 
neutral-neutral reactions $\rm H(OH,H_2)O$; $\rm H_2(OH,H)H_2O$; 
and $\rm H_2O(H,H_2)OH$

\subsection{Example results for a typical diffuse molecular cloud}

\begin{figure}
\includegraphics[scale=0.7]{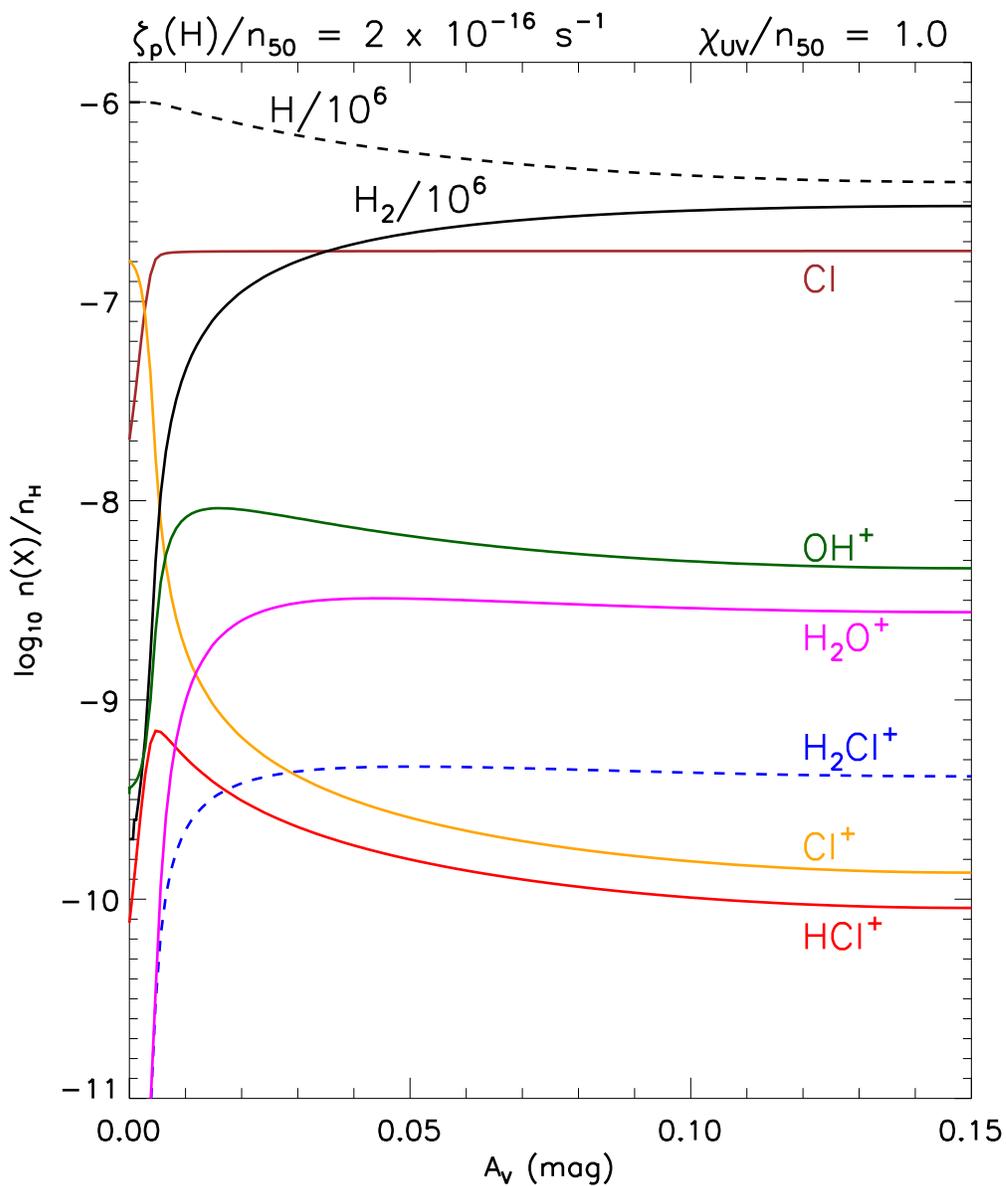}
\caption{Abundance profiles in an interstellar cloud with properties
typical of the diffuse ISM: $\zeta_p({\rm H})/n_{50} =2 \times 10^{-16}\,\rm s^{-1}$, 
$\chi_{UV}/n_{50} = 1$, and $A_{\rm V}({\rm tot})=0.3$.  $A_{\rm V}$ represents the extinction from 
left edge of the cloud, and cloud center is represented by the right edge of the figure.} 
\end{figure}

In Figure 4, we show the abundance profiles predicted for several species within
a diffuse cloud with environmental conditions typical of the diffuse ISM:
$\zeta_p({\rm H})/n_{50} = 2 \times 10^{-16}\,\rm s^{-1}$ and $\chi_{UV}/n_{50} = 1$.
These results plotted here were obtained for a cloud of total visual extinction 
$A_V({\rm tot})=0.3\,\rm mag$, and show the predicted abundances for several 
species as a function of position in the cloud.  Here, the horizontal axis shows 
the distance from the cloud surface, measured in magnitudes visual extinction, 
$A_{\rm V}$, and the vertical axis shows the logarithm of the abundances relative to H nuclei. 
Thus the left edge of the plot represents the irradiated surface of the cloud and the
right edge represents its center.   

The H$_2$ abundance rises sharply near the
cloud edge, owing to the effects of self-shielding in the dipole-allowed Lyman and
Werner band transitions that can lead to photodissociation.  This sharp increase is
tracked closely by the abundances of OH$^+$ and HCl$^+$, which are formed in exothermic
reactions of H$_2$ with O$^+$ and Cl$^+$.  As the H$_2$ fraction increases further, 
significant abundances of $\rm H_2O^+$ and $\rm H_2Cl^+$ appear, the result of further
exothermic reactions of H$_2$ with OH$^+$ and HCl$^+$.  Whereas oxygen is primarily
neutral at the cloud surface, chlorine is mainly ionized because its ionization 
threshold lies (slightly) longward of the Lyman limit.  Once the H$_2$ fraction
exceeds $\sim 10^{-3}$, however, chlorine becomes primarily neutral, owing
to the reaction of Cl$^+$ with H$_2$ to form HCl$^+$.   \bl{Although the abundances
of the oxygen- and chlorine-bearing species are shown in Figure 4 with $A_V$ as
the independent variable, the fundamental quantity that controls their abundances in
this regime is the molecular fraction $2n({\rm H}_2)/n_{\rm H}$.}    

\begin{figure}
\includegraphics[scale=0.65]{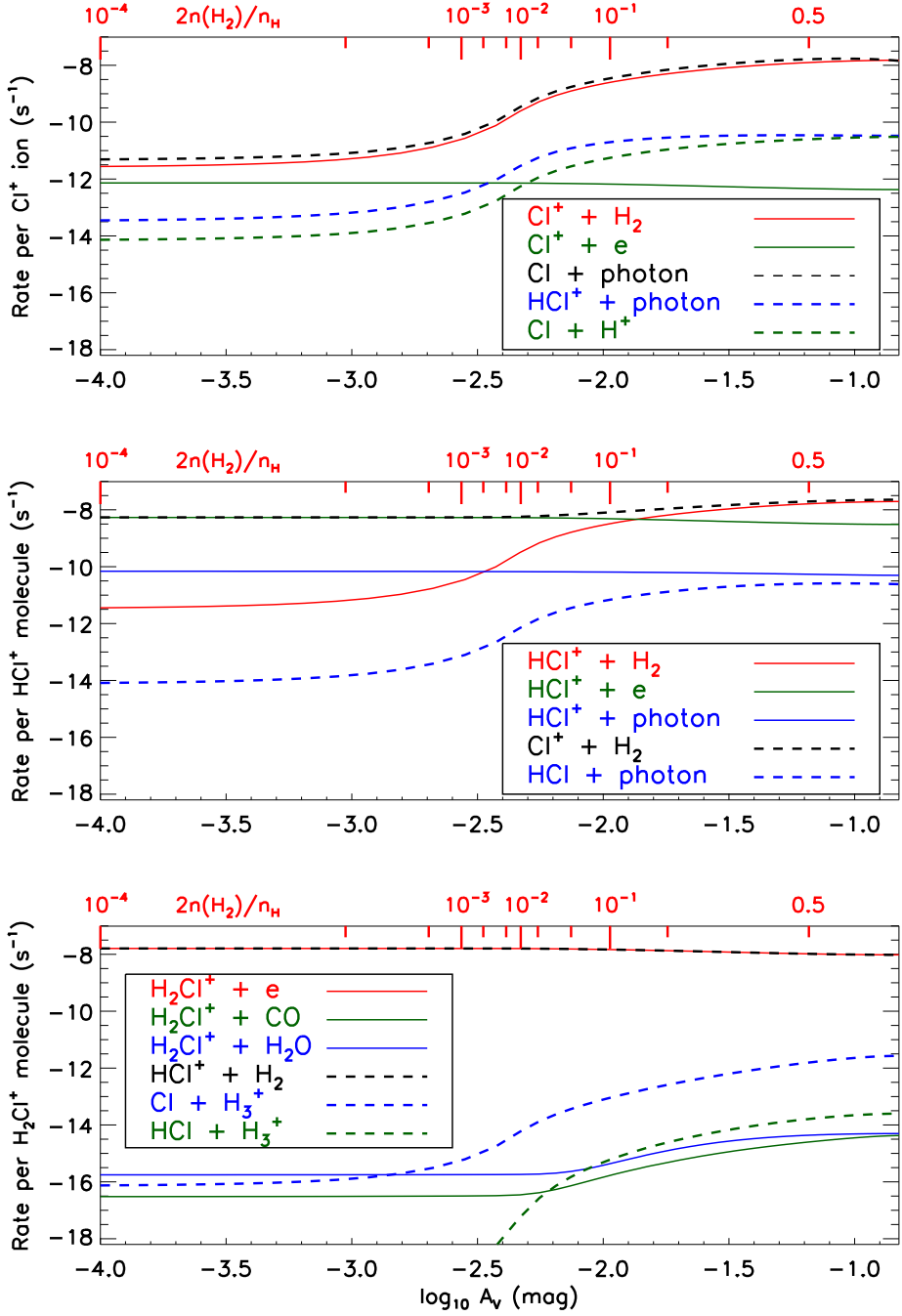}
\caption{Rates of formation (dashed curves) and destruction (solid curves) per particle
for Cl$^+$ (top), HCl$^+$ (middle), and H$_2$Cl$^+$ (bottom). 
$A_{\rm V}$, plotted here on a logarithmic scale, represents the extinction from 
left edge of the cloud, and cloud center is represented by the right edge of the figure. 
Results are shown for $\zeta_p({\rm H})/n_{50} =2 \times 10^{-16}\,\rm s^{-1}$, 
$\chi_{UV}/n_{50} = 1$, and $A_{\rm V}({\rm tot})=0.3$.} 
\end{figure}

Figure 5 shows the rates of various reactions in the chemical network.  Here,
the rates of various destruction and formation processes (solid and dashed curves, 
respectively) are shown for the three key chlorine-bearing ions: Cl$^+$ (top panel), 
HCl$^+$ (middle panel), and H$_2$Cl$^+$.  The distance from the 
cloud surface, again measured in magnitudes of visual extinction, is shown on a 
logarithmic scale for clarity.  Red tick marks at the top of each panel
indicate the local molecular fraction, $2n({\rm H}_2)/n_{\rm H}$.
Figure 5 indicates that for
the example cloud parameters considered here, the chemistry of these
ions is controlled by a relatively small set of processes.  Photoionization of Cl
dominates the formation of Cl$^+$ throughout the cloud, while reaction with H$_2$
dominates its destruction everywhere.  Near the cloud surface, radiative recombination
makes a significant additional contribution to the destruction of Cl$^+$.  The formation 
of HCl$^+$ is  dominated by the reaction of Cl$^+$ with H$_2$,
and its destruction is dominated by DR (near the surface where $2n({\rm H}_2)/n_{\rm H}
\simlt 0.1$) or reaction with H$_2$ (in
the cloud interior).  The latter reaction completely dominates the formation of H$_2$Cl$^+$,
for which the only significant destruction process is DR.  Beyond the six processes
mentioned above -- comprising two DR reactions, two hydrogen abstraction reactions, and the
photoionization and radiative recombination of Cl -- no other process contributes at a level
of more than $\sim 2\%$ to the total formation or destruction rate
in the case shown here.

\section{Discussion}

The abundance profiles shown in Figure 4 apply to a single cloud model with given 
values of $\zeta_p({\rm H})/n_{50}$,
$\chi_{UV}$, and $A_{\rm V}({\rm tot})$.
For this case, and for every other cloud model in the grid we constructed, we
may compute the column densities of each species Y, $N({\rm Y})$, for comparison with the
measured values.

\subsection{Dependence of the $N({\rm H_2Cl^+})/N({\rm HCl^+})$  and 
$N({\rm H_2O^+})/N({\rm OH^+})$ \\ ratios on the molecular fraction}

The chemical network shown in the Figure 3 suggests that the 
$N({\rm H_2Cl^+})/N({\rm HCl^+})$ ratio, like the $N({\rm H_2O^+})/N({\rm OH^+})$
ratio (e.g. Neufeld et al.\ 2010), is controlled by the molecular fraction in the ISM.
This is demonstrated in Figure 6, where the $N({\rm H_2Cl^+})/N({\rm HCl^+})$ 
and $N({\rm H_2O^+})/N({\rm OH^+})$ ratios (shown in blue and magenta, respectively)
are plotted as a function of $A_{\rm V}({\rm tot})$.  All the results shown here
were obtained for the same environmental parameters, 
$\zeta_p({\rm H})/n_{50} = 2 \times 10^{-16}\,\rm s^{-1}$ and $\chi_{UV}/n_{50} = 1$.  
Red tickmarks on the top axis indicate how the average
molecular fraction, $2N({\rm H}_2)/N_{\rm H}$, varies with total visual extinction;
and dashed horizontal lines indicate the column density ratios
observed  for the entire [17,80]~km/s
velocity interval covered by well-separated foreground gas along the W49N sight-line.
As expected, the observed $N({\rm H_2X^+})/N({\rm HX^+})$ ratio generally increases
with the molecular fraction for both elements (X = O and X = Cl)
\footnote{\bl{For very low values of the molecular fraction, 
the $N({\rm H_2O^+})/N({\rm OH^+})$
ratio exhibits a floor at around 0.01.  In this limit, which is not
of relevance to the observations reported here, the model predicts
that OH$^+$ and H$_2$O$^+$ would be formed 
by the photoionization of OH or H$_2$O molecules produced on 
(and then photodesorbed from) grain surfaces.}}
However, when compared with the observed ratios, the standard reaction rates 
(dashed blue curve) yield a clear discrepancy between
the predictions for $N({\rm H_2O^+})/N({\rm OH^+})$ (magenta) and those for 
$N({\rm H_2Cl^+})/N({\rm HCl^+})$.
A model with the $2N({\rm H}_2)/N_{\rm H}$ value $\sim 0.1$ (corresponding to 
$A_{\rm V}({\rm tot}) \sim 0.05$ for these environmental parameters) underpredicts
the $N({\rm H_2Cl^+})/N({\rm HCl^+})$ ratio by almost an order-of-magnitude.
\bl{Moreover, there is no value $A_{\rm V}({\rm tot})$ for which 
the predicted $N({\rm H_2Cl^+})/N({\rm HCl^+})$ ratio is more than one-third the
value observed.}

\begin{figure}
\includegraphics[scale=0.7]{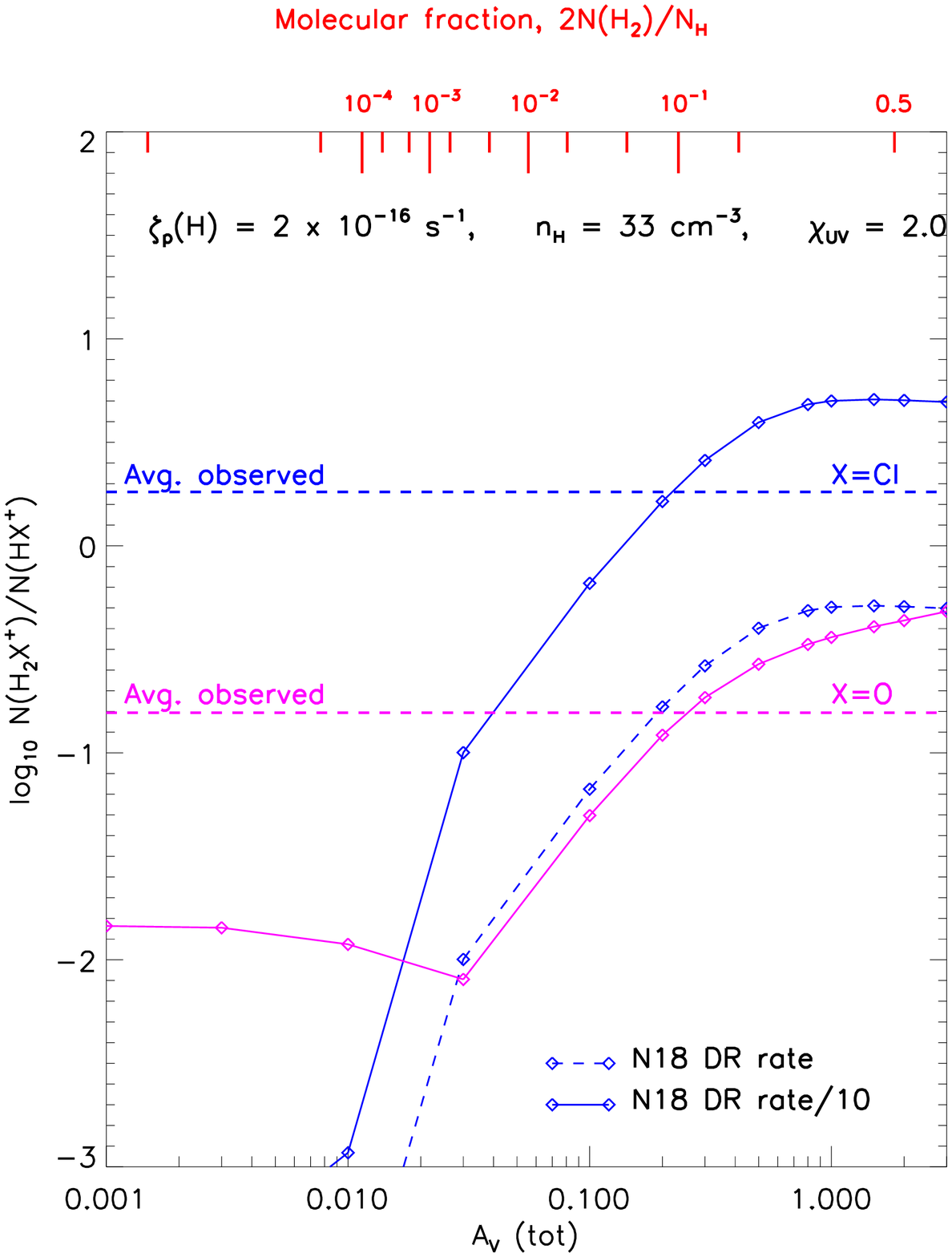}
\caption{Predicted column density ratios, $N({\rm H_2Cl^+})/N({\rm HCl^+}$ (blue) and 
$N({\rm H_2O^+})/N({\rm OH^+}$} (magenta), as a function of the total 
visual extinction through a cloud,$A_{\rm V}({\rm tot})=0.3$.  Results are shown for 
$\zeta_p({\rm H})/n_{50} =2 \times 10^{-16}\,\rm s^{-1}$ and $\chi_{UV}/n_{50} = 1$.
Horizontal dashed lines show the values observed for the entire [17,80]~km/s
velocity interval covered by well-separated foreground gas along the W49N sight-line.
Dashed blue curve: predictions with the ${\rm H_2Cl^+}$ DR rate obtained by N18; 
solid blue curve: predictions with one-tenth that DR rate (see the text).
\end{figure}

The solid blue curve shows $N({\rm H_2Cl^+})/N({\rm HCl^+})$ predictions we obtain
with the assumed rate for H$_2$Cl$^+$ DR reduced by a factor 10.  Not surprisingly, since
DR dominates the destruction of ${\rm H_2Cl^+}$, the solid blue curve lies 
a factor of 10 above the dashed blue curve and removes the discrepancy described above.
Unless some important ${\rm H_2Cl^+}$ formation mechanism has been overlooked 
in our analysis, the observations argue strongly for an H$_2$Cl$^+$ DR rate that is 
an order-of-magnitude smaller than that adopted in our standard model.  The rate
coefficient required to match the observed $N({\rm H_2Cl^+})/N({\rm HCl^+})$ ratio 
is in fact entirely consistent with the measurement of Kawaguchi et al.\ (2016).
The preponderance of the experimental evidence, however, points to a considerably 
larger DR rate (Geppert et al.\ 2009, W16, N18) than that measured by 
Kawaguchi et al.\ (2016), so the observed $N({\rm H_2Cl^+})/N({\rm HCl^+})$ ratio
remains a puzzle.  Solutions to this puzzle include the possibility that the
DR rate for $N({\rm H_2Cl^+})$ is anomalously low in its ground rotational state
that is predominantly populated at the low density of diffuse interstellar clouds.

\subsection{Column densities of molecular ions relative to those of atomic 
hydrogen}

In addition to comparing the observed and predicted values of the ratios
$N({\rm H_2Cl^+})/N({\rm HCl^+})$ and $N({\rm H_2O^+})/N({\rm OH^+})$,
it is also valuable to consider the column densities of all four molecular
ions relative to that of atomic hydrogen.  In this part of the analysis, we have fine-tuned the environmental parameters to
optimize the fit to $N({\rm HCl^+})/N({\rm H})$, $N({\rm OH^+})/N({\rm H})$ and 
$N({\rm H_2O^+})/N({\rm H})$.  Here, we interpolated between grid points and allowed the
values of the environmental parameters  $\zeta_p({\rm H})/n_{50}$ 
and $\chi_{UV}/n_{50}$ to vary between one-third and three times the ``typical"
diffuse cloud parameters adopted for Figures 4 -- 6.  Taking the typical density
as $33\,\rm cm^{-3}$ (i.e. $n_{\rm 50}=0.66$) following Wolfire et al.\ (2003)
\footnote{
For foreground material in the $50-80 \, \rm km\,s^{-1}$ range,  
this value is consistent with the density inferred by Gerin et al.\ (2015; hereafter G15) 
from observations of C$^+$ ($n_{\rm H} = 38 \pm 5 \rm cm^{-3}$), whereas for material in the
$25-50 \, \rm km\,s^{-1}$ range, G15 inferred a density that was a factor of 
$\sim 2.5$ larger than our adopted value.  Since the  molecular column densities predicted by
the model are a function of $\zeta_p({\rm H})/n_{50}$ 
and $\chi_{UV}/n_{50}$ (see NW17), the best-fit parameters given here may be
easily scaled for different preferred values of $n_{\rm H}$.},
we were able to obtain the best simultaneous 
fit to $N({\rm HCl^+})/N({\rm H})$, $N({\rm OH^+})/N({\rm H})$ and 
$N({\rm H_2O^+})/N({\rm H})$ with an assumed $\zeta_p({\rm H})$ of $2.0 \times 10^{-16}
\,\rm s^{-1}$ and an assumed $\chi_{UV}$ of 2.0.  Given these environmental 
parameters and an assumed total visual extinction of 0.2~mag, we were
able to fit the ${\rm HCl^+}$, ${\rm OH^+}$, and ${\rm H_2O^+}$ 
column densities to better than $20\%$.  As implied by Section 4.2 above,
the ${\rm H_2Cl^+}$ abundance was, of course, underpredicted by an order-of-magnitude
given the standard DR given in Table 2; with the assumed ${\rm H_2Cl^+}$ DR rate reduced by
a factor 10, the column densities of all four molecular ions, relative to that of atomic hydrogen,
could be fit simultaneously for entirely reasonable environmental parameters.

\begin{figure}
\includegraphics[scale=0.7]{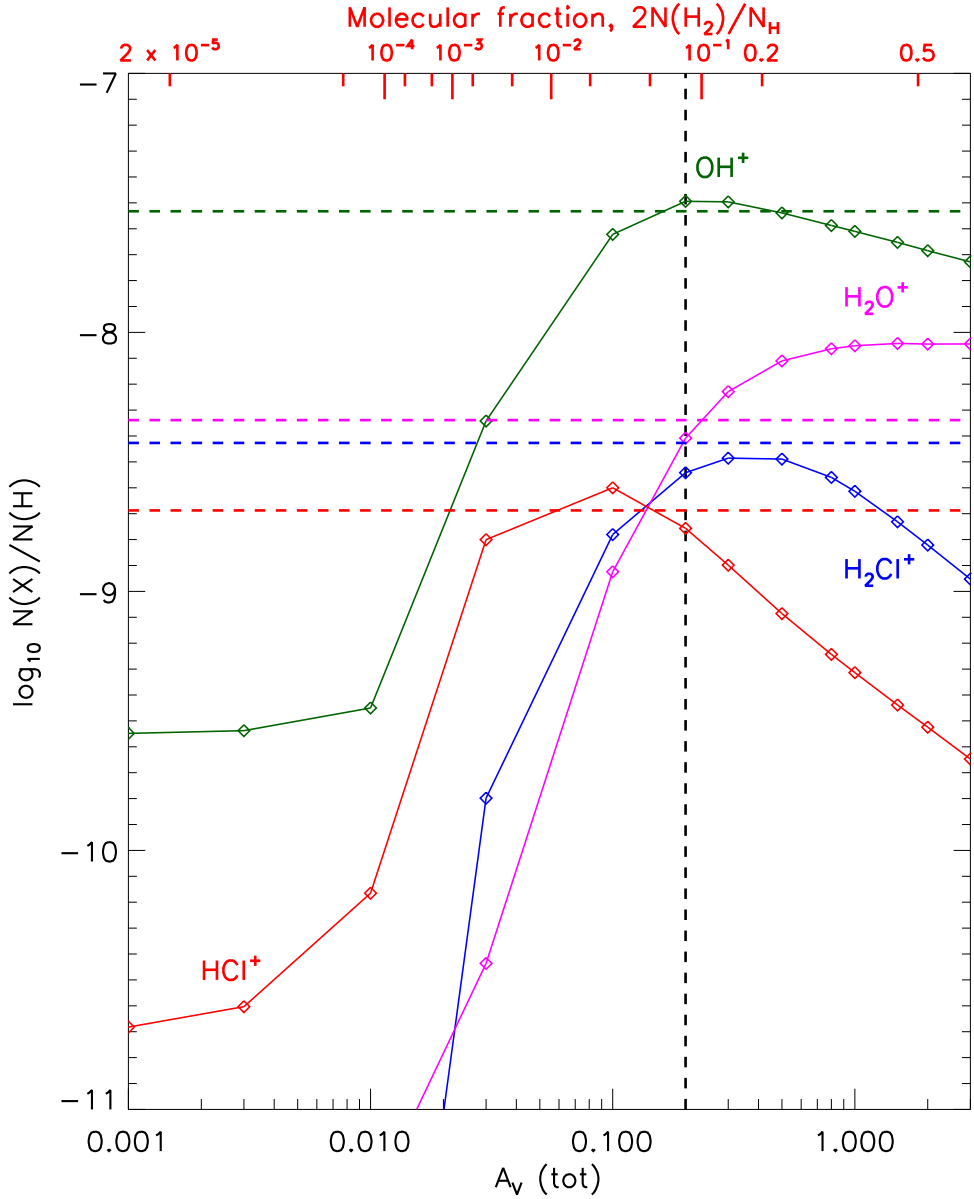}
\caption{Predicted abundances relative to atomic hydrogen
for the optimized environmental parameters 
$\zeta_p({\rm H}) =2 \times 10^{-16}\,\rm s^{-1}$, $n_{\rm H}=33\, \rm cm^{-3}$, 
and $\chi_{UV} = 2$.  The four molecular ion abundances 
are fit to within $20\%$ for $A_{\rm V}({\rm tot})=0.2$.  Here, we adopted 
an ${\rm H_2Cl^+}$ DR rate one-tenth that obtained by N18.}
\end{figure}

The column density ratios $N({\rm HCl^+})/N({\rm H})$, $N({\rm H_2Cl^+})/N({\rm H})$, 
$N({\rm OH^+})/N({\rm H})$ and $N({\rm H_2O^+})/N({\rm H})$ are shown in Figure 7
as a function of $A_{\rm V}({\rm tot})$.  These results apply for  
the optimal environmental parameters $\zeta_p({\rm H})=2.0 \times 10^{-16}
\,\rm s^{-1}$, $n_{\rm H} = 33\,\rm cm^{-3}$ and  $\chi_{UV}=2.0$.  
The dashed horizontal lines indicate the values observed for the entire [17,80]~km/s
velocity interval covered by well-separated foreground gas along the W49N sight-line;
here, the color-coding follows that adopted the theoretical predictions, with results for 
$N({\rm HCl^+})/N({\rm H})$, $N({\rm H_2Cl^+})/N({\rm H})$,
$N({\rm OH^+})/N({\rm H})$ and $N({\rm H_2O^+})/N({\rm H})$ shown respectively 
in red, blue, green and magenta.  The dashed vertical line indicates where the
fit is optimized at a \bl{molecular fraction $\sim 8\%$ (achieved for 
$A_{\rm V}({\rm tot}) \sim 0.2$~mag)}.

The satisfactory fit to the $N({\rm HCl^+})/N({\rm H})$ described above contrasts with
the conclusion reached by DL12 that the observed abundance
of ${\rm HCl^+}$ exceeded the NW09 model predictions by a factor of 3.  Three
effects contribute to this difference: (1) the $N({\rm HCl^+})$ obtained in the
present study is almost $40\%$ smaller than that inferred from the earlier {\it Herschel} 
spectrum; (2) at the kinetic temperature $\sim 120$~K predicted
in the gas responsible for the ${\rm HCl^+}$ absorption, the ${\rm HCl^+}$ DR rate 
adopted here is $\sim 40\%$ smaller than that adopted by NW09; (3) the optimum
fit reported here is obtained for an enhanced (but not unreasonably high) UV field
twice the value considered in the models of NW09. 

\subsection{Variations along the sight-line}

\begin{figure}
\includegraphics[scale=0.7]{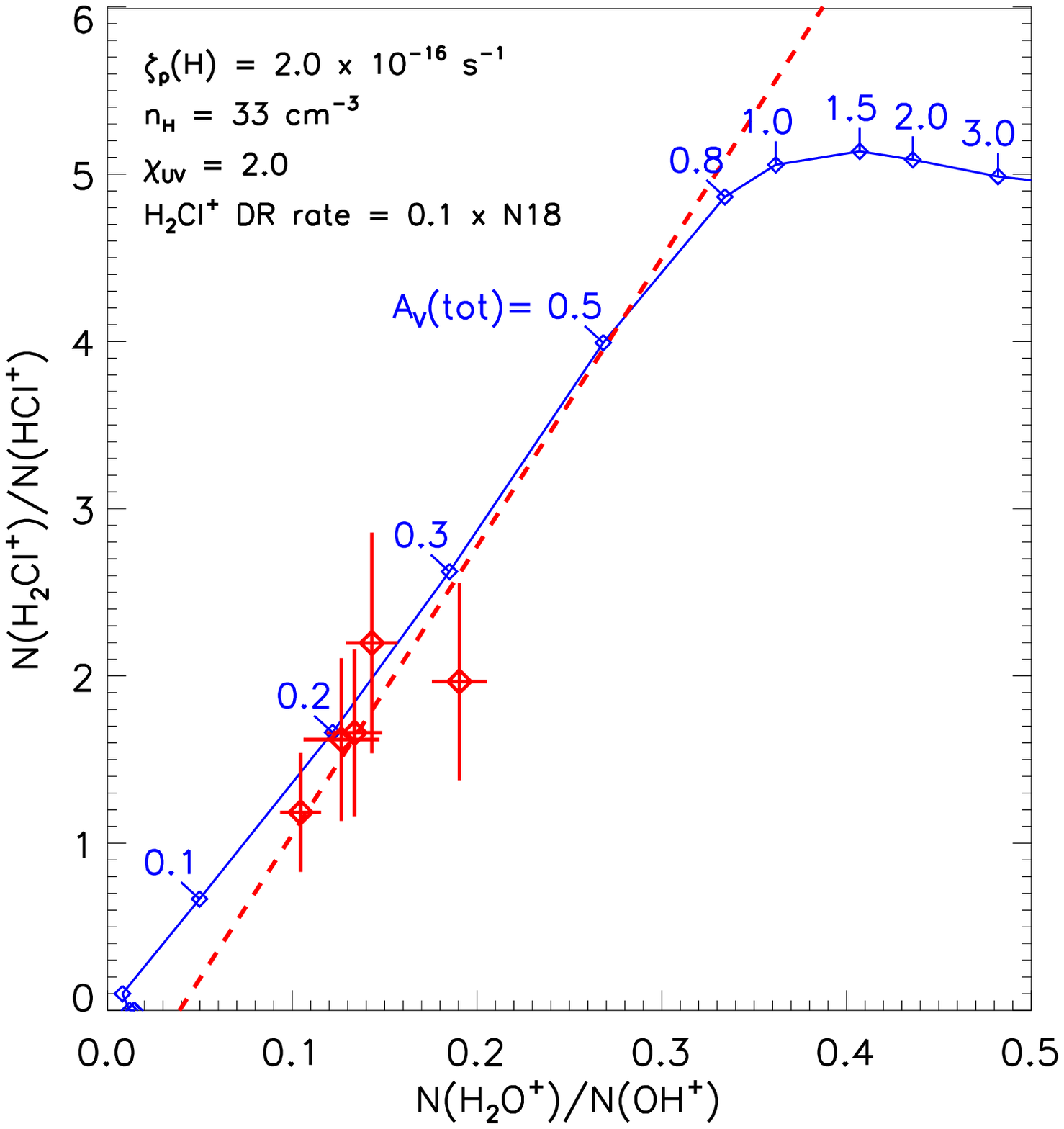}
\caption{Predicted column density ratios, $N({\rm H_2Cl^+})/N({\rm HCl^+}$ (vertical axis) 
and $N({\rm H_2O^+})/N({\rm OH^+}$ (horizontal axis), for the optimized environmental parameters.  
Here, we adopted 
an ${\rm H_2Cl^+}$ DR rate one-tenth that obtained by N18.
Red points show  the observed ratios for the five velocity intervals in which they can be determined 
reliably.  The dashed red line shows the best linear fit to the observed data
(with a slope that has a 1 $\sigma$ fractional uncertainty of $\sim 55\%$).}
\end{figure}

Finally, in Figure 8, we present (red points) the observed
$N({\rm H_2Cl^+})/N({\rm HCl^+})$ and $N({\rm H_2O^+})/N({\rm OH^+})$ ratios 
for the five velocity intervals in Table 2 that are well separated from 
the source velocity.   These are compared with the theoretical predictions
obtained for the optimal environmental parameters ($\zeta_p({\rm H})=2.0 \times 10^{-16}
\,\rm s^{-1}$, $n_{\rm H} = 33\,\rm cm^{-3}$ and $\chi_{UV}=2.0$)
and with the ${\rm H_2Cl^+}$ DR rate a factor 10 below the standard value.
The predictions for this model are represented by the blue locus with
values of $A_{\rm V}({\rm tot})$ marked in blue. 

As expected the observed $N({\rm H_2Cl^+})/N({\rm HCl^+})$ 
and $N({\rm H_2O^+})/N({\rm OH^+})$  ratios are positively-correlated 
with a Pearson correlation coefficient of 0.7 and a best fit linear regression 
represented by the dashed red line.  However, the measured correlation, although
suggestive, does not quite reach even a 2 $\sigma$ level of statistical significance.

\section{Summary}

\noindent (1) We have used the GREAT instrument on SOFIA to observe the 
$^2\Pi_{3/2}\, J = 5/2 – 3/2$ transition of HCl$^+$ near 1444~GHz 
toward the bright THz continuum source W49N.

\noindent (2) The resultant HCl$^+$ spectrum reveals absorption by diffuse
foreground gas unassociated with the background continuum source.

\noindent (3) A comparison with previous {\it Herschel} observations of H$_2$Cl$^+$
suggests that the $n({\rm H_2Cl^+})/n({\rm HCl^+})$ abundance ratio
varies by a factor of at most 2 along the sight-line.

\noindent (4) We have constructed a grid of diffuse cloud models in which the
equilibrium abundances of ${\rm HCl^+}$ and ${\rm H_2Cl^+}$ are
computed as function of
the cloud properties.  These models incorporate updates to the
rates adopted by NW09 for several significant reactions.

\noindent (5) The total column density ratio,
 $N({\rm H_2Cl^+})/N({\rm HCl^+})$, within the diffuse foreground gas 
is an order-of-magnitude larger than the predictions of our standard 
diffuse cloud model.  In that model, we adopted a dissociative recombination rate for
${\rm H_2Cl^+}$ that reflects the values typically obtained in recent laboratory
measurements.  This discrepant $N({\rm H_2Cl^+})/N({\rm HCl^+})$ ratio suggests that
recent laboratory values for the ${\rm H_2Cl^+}$ or ${\rm D_2Cl^+}$ DR rates
are inapplicable, perhaps because the laboratory studies apply to rotationally-warm 
molecular ions whereas ${\rm H_2Cl^+}$ is 
rotationally-cold in low density interstellar clouds. 

\noindent (6)  For the molecular ions ${\rm HCl^+}$, ${\rm OH^+}$, and ${\rm H_2O^+}$,
the model predictions can provide a satisfactory fit to the observed 
column densities along the  W49N sight-line.  For a cloud density typical of the diffuse ISM, 
$n_{\rm H} = 33\, \rm cm^{-3}$, the optimal parameters are
an assumed cosmic-ray ionization rate of 
$\zeta_p({\rm H}) = 2 \times 10^{-16}\,\rm s^{-1}$, an interstellar radiation field 
$\chi_{\rm UV}$ of 2.0 (i.e. twice the standard value), and a total visual 
extinction $A_V({\rm tot}) = 0.2$~mag across an individual cloud \bl{(leading to a
molecular fraction $2N({\rm H}_2)/N_{\rm H} \sim 8\%$).}

\begin{acknowledgements}

\bl{Based on observations made with the NASA/DLR Stratospheric Observatory for Infrared Astronomy.
SOFIA Science Mission Operations are conducted jointly by the Universities Space Research Association, Inc., 
under NASA contract NAS2-97001, and the Deutsches SOFIA Institut under DLR contract 50 OK 0901.   
This research was supported by USRA through a grant for SOFIA Program 06-0017.  
H.G. acknowledges support from the National Science 
Foundation for participation in this work while serving there. Any opinions, 
findings, and conclusions expressed in this material are those of the authors 
and do not necessarily reflect the views of the National Science Foundation.
It is a pleasure to acknowledge helpful discussions
with Daniel Savin, Oldrich Novotn{\'y}, and Andreas Wolf.
We gratefully acknowledge the outstanding support provided by the SOFIA Operations Team 
and the GREAT Instrument Team.} 

\end{acknowledgements}

\end{document}